\documentstyle[multicol,aps,epsf]{revtex}   
\begin{document}   
   
\title{Operator-algebraic approach to the yrast spectrum of   
weakly interacting trapped bosons    
}\par   
   
\author{Masahito Ueda$^1$ and Tatsuya Nakajima$^{2, \ast}$}\par   
   
\address{$^1$ Department of Physics, Tokyo Institute of Technology,   
Meguro-ku, Tokyo 152-8551, Japan, and \\   
CREST, Japan Science and Technology Corporation (JST), Saitama 332-0012,   
Japan   
}\par   
   
\address{$^2$ Department of Physics, The University of Texas,    
Austin, Texas 78712, USA   
}\par   
   
%\date{\today}   
\date{August 16, 2001}   
   
\maketitle   
   
\begin{abstract}   
We present an operator-algebraic approach to deriving the low-lying   
quasi-degenerate spectrum of weakly interacting trapped $N$ bosons    
with total angular momentum $\hbar L$ for the case of $L/N\ll 1$,     
demonstrating that the lowest-lying excitation spectrum is given by   
$27 g n_3(n_3-1)/34$, where $g$ is the strength of the repulsive      
contact interaction and $n_3$ the number of excited octupole quanta.  
Our method provides constraints for these quasi-degenerate many-body  
states and gives higher excitation energies that depend linearly on   
$N$.    
\end{abstract}   
   
   
\begin{multicols}{2} 
\narrowtext   
\section{Introduction}   
   
Realization of Bose-Einstein condensates (BECs) in atomic gases              
~\cite{Anderson,Davis,Bradley} has stimulated explosively growing interest   
in several subfields of physics.    
With unprecedented controllability in creating, manipulating, and probing    
systems, many previously ``gedanken" experiments have become or are          
expected to become a reality~\cite{review}.   
Another characteristic feature of gaseous BEC systems is their good          
isolation from the surrounding environment, as the system is suspended in    
a vacuum chamber by a magnetic or an optical trapping potential.             
When the potential is axisymmetric, the projected angular momentum (AM) on   
the symmetry axis is conserved and can take on any integral multiple of      
$\hbar$.    
In contrast, in bulk superfluid systems, the container plays the role of a   
highly dissipative environment so that the system can sustain AM only at     
thermodynamically stable values, resulting in the quantization of            
circulation.    
In this respect, gaseous BEC systems are more like nuclear systems than      
bulk superfluid systems, and it is of interest to study the energy           
spectrum of a weakly interacting BEC system as a function of AM, namely      
the yrast spectrum~\cite{wilk,thomas,Linn,mot}.   
   
The lowest-energy state of an isolated system subject to a given AM,         
$\hbar L$, is called the yrast state~\cite{Hamamoto}.                        
When the system is in the yrast state, all kinetic energy is used for        
rigid-body rotation, so the system is at zero temperature. This allows       
one to measure fine energy levels ---the yrast spectrum--- close to the      
ground state.    
Recently, the yrast state has been extensively discussed in the context      
of gaseous BEC systems~\cite{wilk,mot,kavou,jackson,bert,jk,sw,pb,NU}.       
Mottelson has pointed out that for repulsive contact interactions the yrast  
spectrum is dominated by quasi-degenerate quadrupole and octupole            
excitations~\cite{mot}, and Kavoulakis {\it et al} have shown that the       
interaction between octupole excitations makes the excitation energy of      
this mode slightly higher than that of the quadrupole mode~\cite{kavou}.     
Bertsch and Papenbrock~\cite{bert} have found numerically that for           
$0\leq L\leq N$, the energy of the yrast state is given by                   
$E=L\hbar\omega+gN(N-1-L/2)$, where $\omega$ is the frequency of the         
confining potential, $g$ the strength of the interaction, and $N$ the        
number of bosons, which we shall assume to be constant.                      
The corresponding eigenstate has been shown to exist in Refs.~\cite{jk,sw,pb}.  
Recently, we have found numerically that the {\it lowest-lying} yrast        
spectrum arises from the pairwise repulsive interaction between octupole     
excitations and that the corresponding interaction energy is given by        
$1.59\,gn_3(n_3-1)/2$, where $n_3$ is the number of octupole   
excitations~\cite{NU}.   
In this paper, we present an operator-algebraic approach~\cite{note} to      
analytically deriving these results in a systematic way, demonstrating that  
the interaction energy between octupole excitations is given by              
$27g n_3(n_3-1)/34$,   
where the coefficient 27/34 agrees precisely with the numerically obtained   
value~\cite{NU}.   
Our method also provides constraints for the quasi-degenerate many-body      
states and gives higher excitation energies that depend linearly on $N$.     
   
This paper is organized as follows.   
Section~\ref{sec:Formulation} describes the model under consideration and    
formulates the problem.   
Section~\ref{sec:Collective} describes collective modes and discusses   
why they behave as well-defined quasiparticles when the system undergoes    
the Bose-Einstein condensation.   
Section~\ref{sec:yrast} derives the yrast spectrum based on our    
operator-algebraic approach, and    
Sec.~\ref{sec:Conclusions} concludes this paper.

%%%%%%%%%%%%%%%%%%%%%%%%%%%%%%%%%%%%%%%%   
\section{Formulation of the Problem}   
\label{sec:Formulation}   
   
We consider a two-dimensional system of harmonically trapped $N$ bosons   
interacting via a contact $\delta$-function interaction.   
The Hamiltonian of the system is given by   
\begin{eqnarray}   
H &=& \sum_{i=1}^N    
\left[- \frac {\hbar ^2}{2M}   
\left (\frac {\partial ^2 \ }{\partial x_i^2}   
+\frac {\partial ^2 \ }{\partial y_i^2} \right )   
+ \frac {M \omega ^2}{2} (x_i^2+y_i^2)\right] \nonumber \\  
& & \qquad \qquad \quad +\frac{2 \pi \hbar  g}{M\omega}   
\sum _{i\neq j}   
\delta ^{(2)} ({\bf r}_i - {\bf r}_j) ,   
\label{H1}   
\end{eqnarray}   
where $M$ is the mass of the boson, $\omega$ the frequency of the      
confining potential, and $g$ the strength of the interaction.     
In the second-quantized form, Eq.~(\ref{H1}) is written as   
\begin{eqnarray}   
\hat{H} &=& \int d^2z\left [   
\hat{\Psi}^\dagger(z)h(z)\hat{\Psi}(z)   
+ \frac{2 \pi \hbar g}{M\omega}   
\hat{\Psi}^\dagger(z)\hat{\Psi}^\dagger(z)   
\hat{\Psi}(z)\hat{\Psi}(z)\right ], \nonumber \\  
& & \label{H3}   
\end{eqnarray}   
where   
$z\equiv x+iy$ and    
\begin{eqnarray}   
h(z) \equiv   
- \frac {2\hbar ^2}{M}   
\frac {\partial ^2 \ \ }{\partial z \partial z^{\ast}}   
+ \frac {M \omega ^2}{2} |z|^2   
\label{h}   
\end{eqnarray}   
denotes the single-particle Hamiltonian.   
   
Throughout this paper we consider the case of very weak interactions,    
{\it i.e.}, $|g| N \ll \hbar\omega$. It is then convenient to take the   
eigenfunctions of $h(z)$ as a basis set for expansion of the field       
operator. We introduce two sets of creation and annihilation operators,  
\begin{eqnarray}   
& & \hat{a}_+=\frac{z^*}{2d}+d\frac{\partial}{\partial z}, \ \ \ \   
 \hat{a}_+^\dagger=\frac{z}{2d}-d\frac{\partial}{\partial z^*},   
\label{a+} \\   
& & \hat{a}_-=\frac{z}{2d}+d\frac{\partial}{\partial z^*}, \ \ \    
 \hat{a}_-^\dagger=\frac{z^*}{2d}-d\frac{\partial}{\partial z},   
\label{a-}   
\end{eqnarray}   
where $d\equiv\sqrt{\hbar/M\omega}$ is a characteristic length scale of     
the ground-state wave function of $h(z)$, and $\hat{a}_+$ (or $\hat{a}_-$)  
annihilates a quantum with magnetic quantum number 1 (or -1) and satisfies  
the canonical commutation relations    
\begin{eqnarray}    
[\hat{a}_+,\hat{a}_+^\dagger]=[\hat{a}_-,\hat{a}_-^\dagger]=1, \ \    
[\hat{a}_+,\hat{a}_-]=[\hat{a}_+,\hat{a}_-^\dagger]=0.   
\end{eqnarray}  
In terms of these operators, $h$ is diagonalized as     
\begin{eqnarray}  
h=\hbar\omega(\hat{a}_+^\dagger \hat{a}_+ +\hat{a}_-^\dagger\hat{a}_-+1).  
\label{h2}      
\end{eqnarray}  
The vacuum state $|0\rangle$ is defined by    
$\hat{a}_+|0\rangle=\hat{a}_-|0\rangle=0$.   
   
The operator corresponding to the magnetic quantum number $m$ is given    
by $\hat{a}_+^\dagger \hat{a}_+-\hat{a}_-^\dagger \hat{a}_-$. Because     
of the symmetry between $\hat{a}_+$ and $\hat{a}_-$, we may consider      
only the case of $m \ge 0$ with no loss of generality. Moreover, when     
$\hbar \omega \gg |g| N$, we may consider only those states               
\{$|\Psi\rangle$\} that satisfy   
\begin{eqnarray}   
\hat{a}_-^\dagger \hat{a}_-|\Psi\rangle=0   
\label{LLL}   
\end{eqnarray}   
because for a given total AM, states that do not satisfy Eq.~(\ref{LLL})  
are at least $2 \hbar \omega$ higher in energy than those satisfying the  
equation.  
Physically, Eq.~(\ref{LLL}) suggests that no particle rotates in a        
clockwise direction. In what follows, we shall restrict ourselves to the  
Hilbert subspace constrained by Eq.~(\ref{LLL}), which corresponds to     
the lowest-Landau-level approximation in the theory of the fractional     
quantum Hall effect.   
   
Within this approximation, a complete set of basis functions can be       
constructed as follows.   
Using $\langle z| \hat{a}_+|0\rangle = 0$ and Eq.~(\ref{a+}), we obtain   
\begin{eqnarray}   
\phi_0(z) \equiv \langle z|0\rangle=\frac{1}{\sqrt{\pi}d}   
e^{-|z|^2/2d^2},   
\label{norm}   
\end{eqnarray}   
where the prefactor is determined so as to satisfy the normalization condition   
$\int d^2z \,|\phi_0(z)|^2=1$.   
The wavefunction $\phi_m(z)$ for the state with magnetic quantum number $m$    
is given by   
\begin{eqnarray}   
\phi_m(z) &\equiv& \langle z|m\rangle   
=\langle z|\frac{(\hat{a}_+^\dagger)^m}{\sqrt{m!}}|0\rangle   
=\frac {1}{\sqrt{m!}}    
\left(\frac{z}{2d}-d\frac{\partial}{\partial z^*}\right)^m   
\langle z|0\rangle \nonumber \\
&=& \frac{z^m}{\sqrt{\pi m!}\,d^{m+1}}e^{-|z|^2/2d^2},   
\label{phi_m}   
\end{eqnarray}   
and satisfies    
the orthonormal conditions    
\begin{eqnarray}   
\int d^2z \,\phi_m^*(z)\phi_n(z)=\delta_{mn}.   
\label{orthonorm}   
\end{eqnarray}   
It follows from Eq.~(\ref{phi_m}) that $| z \rangle$    
is a coherent state defined by   
\begin{eqnarray}   
| z \rangle = (\sqrt{\pi} d)^{-1} \,    
\exp [(z^\ast \hat{a}_{+}^{\dagger} - z \hat{a}_{+})/d] \,| 0 \rangle.   
\label{coh}   
\end{eqnarray}   
Expanding the field operator in terms of $\phi_m(z)$ as   
\begin{eqnarray}   
\hat{\Psi}(z)=\sum_m\hat{b}_m\phi_m(z),   
\label{Psi}   
\end{eqnarray}   
and substituting this into Eq.~(\ref{H3}),    
we obtain   
\begin{eqnarray}   
\hat{H}=\hbar\omega \hat{L}   
+ g\sum_{m_1,\cdots,m_4}   
V_{m_1m_2m_3m_4}   
\hat{b}_{m_1}^\dagger   
\hat{b}_{m_2}^\dagger   
\hat{b}_{m_3}   
\hat{b}_{m_4} ,   
\label{H4}   
\end{eqnarray}   
where    
\begin{eqnarray}    
\hat{L} \equiv \sum_m m\hat{b}_m^\dagger\hat{b}_m    
\label{L}    
\end{eqnarray}    
gives the total magnetic quantum number and    
$V_{m_1m_2m_3m_4}$ is the matrix element of    
the interaction given as~\cite{bert}  
\begin{eqnarray}    
V_{m_1 m_2 m_3 m_4} =    
\frac {\delta _{m_1+m_2, m_3+m_4} \,(m_1+m_2)!}    
{2^{m_1+m_2}\,\sqrt{m_1 !\,m_2 !\,m_3 !\,m_4 !}} .   
\label{V}    
\end{eqnarray}    
In Eq.~(\ref{H4}) the term $\hbar\omega\hat{N}$ is dropped   
because the total number of bosons   
\begin{eqnarray}    
\hat{N} \equiv \sum_m \hat{b}_m^\dagger\hat{b}_m    
\label{N}    
\end{eqnarray}      
is conserved in our system.    
   
Because $\hat{L}$ is also conserved in our system, we shall henceforth    
focus on the second term on the right-hand side (rhs) of Eq.~(\ref{H4}),  
which will be denoted as $\hat{V}=g\hat{U}$. Our primary goal is to       
diagonalize $\hat{U}$ in operator form within the Hilbert subspace of     
non-negative $m$ and under the two constraints   
\begin{equation}   
\hat{L} = L, \ \ \hat{N} = N .   
\label{cnstr}   
\end{equation}

%%%%%%%%%%%%%%%%%%%%%%%%%%%%%%%%   
\section{Collective modes}   
\label{sec:Collective}   
   
Mottelson has pointed out that low-lying excitations from the yrast   
line can be described in terms of collective modes excited by         
$(d^{-\lambda}/\sqrt{N \lambda !}) \sum_{j=1}^N z_j^{\lambda}$,       
where $\lambda$ is a positive integer. In the second-quantized form,  
the creation operators for the collective modes are written as        
\begin{eqnarray}    
\hat{Q}_\lambda^\dagger &=& \int d^2 z \,   
\hat{\Psi}^\dagger (z)\frac{d^{-\lambda}}{\sqrt{N \lambda !}}   
z^\lambda \,\hat{\Psi}(z)   \nonumber \\
&=& \frac{1}{\sqrt{N}}\sum_{m=0}^\infty\sqrt{{}_{\lambda+m}C_\lambda} \    
\hat{b}_{m+\lambda}^\dagger\hat{b}_m ,   
\label{Q}   
\end{eqnarray}   
where ${}_{m}C_n = m!/n! (m-n)!$. The commutation relation between        
$\hat{Q}_\lambda$ and the interaction Hamiltonian $\hat{V}$ is            
calculated to be   
\begin{eqnarray}   
[\hat{V},\hat{Q}_\lambda]   
&=&   
\frac{2g}{\sqrt{N\lambda!}}\sum_{m_1,\cdots,m_4}   
\frac {\delta _{m_1+m_2, m_3+m_4-\lambda} \,(m_1+m_2)!}   
{2^{m_1+m_2+\lambda}\,\sqrt{m_1 !\,m_2 !\,m_3 !\,m_4 !}}    
\nonumber \\   
& & \times    
\left [   
2^\lambda\frac{m_4!}{(m_4-\lambda)!}-   
\frac{(m_1+m_2+\lambda)!}{(m_1+m_2)!}   
\right ]   
\nonumber \\   
& & \qquad \qquad \quad \times
\hat{b}_{m_1}^\dagger   
\hat{b}_{m_2}^\dagger   
\hat{b}_{m_3}   
\hat{b}_{m_4} .   
\label{VQ}   
\end{eqnarray}   
It follows from this relation that except for   
$\lambda=1$, $\hat{Q}_\lambda$       
does not commute with $\hat{V}$. The collective mode excited by           
$\hat{Q}_\lambda$ ($\lambda>1$) is therefore not, in general, a           
well-defined quasiparticle. The only exception is the dipole operator     
$\hat{Q}_1$, for which the rhs of Eq.~(\ref{VQ}) vanishes. The dipole     
mode is therefore a well-defined collective mode, and for any $L$, an     
eigenstate of the Hamiltonian~(\ref{H4}) may be   
constructed as~\cite{wilk,mot}   
\begin{eqnarray}   
|\Psi^L\rangle=(\hat{Q}_1^\dagger)^L|L=0\rangle,   
\label{Psi^L}   
\end{eqnarray}   
where $|L=0\rangle$ denotes the exact ground state of the   
Hamiltonian~(\ref{H4}) with zero total AM, and hence    
$\hat{V}|L=0\rangle=gN(N-1)|L=0\rangle$.   
The corresponding eigenenergy is given by   
\begin{eqnarray}  
E^{\rm dipole}= \hbar \omega L + gN(N-1)  
\end{eqnarray}  
because   
\begin{eqnarray*}   
\hat{H}\,|\Psi^L\rangle&=&(\hbar\omega \hat{L}+\hat{V})\,   
(\hat{Q}_1^\dagger)^L\,|L=0\rangle  \nonumber \\  
&=&\hbar\omega L\,|\Psi^L\rangle   
+(\hat{Q}_1^\dagger)^L\,\hat{V}    
|L=0\rangle    
=E^{\rm dipole}\,|\Psi^L\rangle.   
\end{eqnarray*}   
The state $|\Psi^L\rangle$ is the lowest-energy state of    
the Hamiltonian with AM $L$ for the case of an attractive interaction ($g<0$)    
and the highest-energy state    
for the case of a repulsive interaction ($g>0$).   
   
It should be noted that while $\hat{Q}_\lambda$ and   
$\hat{Q}_\mu$ commute for any $\lambda$ and $\mu$,   
$\hat{Q}_\lambda$ and $\hat{Q}_\mu^\dagger$ do not. In fact,   
\begin{eqnarray}   
[\hat{Q}_\lambda, \hat{Q}_\mu^\dagger]   
&=& \frac{1}{N}   
\sum_{m=0}^\infty    
\left [   
\theta(\lambda-\mu)    
\sqrt{   
{}_{m+\lambda}C_\lambda \    
{}_{m+\lambda}C_\mu   
}   
\, \hat{b}_m^\dagger \hat{b}_{m+\lambda-\mu} \right.   
\nonumber \\   
& & \left.   
+\theta(\mu-\lambda)    
\sqrt{   
{}_{m+\mu}C_\lambda \ {}_{m+\mu}C_\mu   
}   
\, \hat{b}_{m+\mu-\lambda}^\dagger \hat{b}_{m} \right. \nonumber \\
& & \left. \qquad 
-\sqrt{   
{}_{m+\lambda}C_\lambda \ {}_{m+\mu}C_\mu   
}   
\, \hat{b}_{m+\mu}^\dagger \hat{b}_{m+\lambda}    
\right ] ,   
\label{commu1}   
\end{eqnarray}   
where $\theta(x)$ is the Heaviside step function, which   
takes on 0, 1/2 and   
1 for $x<0$, $x=0$ and $x>0$, respectively.   
The commutation relation~(\ref{commu1}) implies that the    
collective modes carrying different angular momenta do not, in general,   
behave independently.   
For $\mu=\lambda$, Eq.~(\ref{commu1}) reduces to   
\begin{eqnarray}   
[\hat{Q}_\lambda,\hat{Q}_\lambda^\dagger]   
&=&   
\frac{1}{N}   
\sum _{m=0}^\infty    
({}_{m+\lambda}C_\lambda   
-{}_mC_\lambda)   
\hat{b}_m^\dagger\hat{b}_m     
\nonumber \\   
&=&    
\cases{   
1                                                                  
   &for $\lambda=1$;\cr   
1+\frac{2L}{N}                                                     
   &for $\lambda=2$;\cr   
1+\frac{3L}{2N}+\frac{3}{2N}   
{\displaystyle \sum_m m^2 \,\hat{b}_m^\dagger\hat{b}_m}   
   &for $\lambda=3$,   
}   
\label{commu2}   
\end{eqnarray}   
where constraints~(\ref{cnstr}) are used    
in deriving the second equality.    
The commutator $[\hat{Q}_\lambda,\hat{Q}_\lambda^\dagger]$   
in Eq.~(\ref{commu2})    
gives constants for $\lambda=1,2$,   
indicating that the dipole ($\lambda=1$) and quadrupole    
($\lambda=2$) modes behave like bosons.   
In contrast, the commutator for the octupole ($\lambda=3$) mode does      
not give a constant, indicating the anharmonicity of this mode. This      
anharmonicity leads to the interaction between octupole excitations,      
as shown in Sec.~\ref{sec:yrast}.   
   
The energy minimization of $\hat{V}$ must be performed subject to         
constraints~(\ref{cnstr}). In the case of $L/N\ll1$, which we shall       
consider below, these constraints require that    
\begin{eqnarray}   
\hat{n}_0={\cal O}(N), \ \ \ \hat{n}_m={\cal O}(N^0) \ \ (m\geq1),   
\label{cond1}   
\end{eqnarray}   
where $\hat{n}_m\equiv\hat{b}_m^\dagger\hat{b}_m$   
and ${\cal O}(N^k)$ denotes terms of the order of or less than $N^k$.    
In this case, $\hat{Q}_\lambda$ may be simplified as    
\begin{eqnarray}   
\hat{Q}_\lambda \simeq \frac{1}{\sqrt{N}}   
\hat{b}_0^\dagger\hat{b}_\lambda    
,   
\label{Q2}   
\end{eqnarray}   
and the commutation relations in Eq.~(\ref{commu1}) reduce to   
\begin{eqnarray}   
[\hat{Q}_\lambda,\hat{Q}_\mu^\dagger]=   
\delta_{\lambda\mu}+{\cal O}(\frac{L}{N}).   
\end{eqnarray}   
The collective modes, therefore, behave as almost independent bosons for  
$L/N\ll1$, with small inter-mode interactions on the order of $L/N$.      
These features are reflected explicitly in the yrast spectrum, as shown   
in the next section.

%%%%%%%%%%%%%%%%%%%%%%%%   
\section{Yrast spectrum}   
\label{sec:yrast}   
   
Our basic idea for the derivation of the yrast spectrum is to rewrite     
the interaction Hamiltonian $\hat{V}=g\hat{U}$ under constraints~(\ref{cnstr})  
so that   
the excitation spectrum manifests itself.   
We first note that $\hat{U}$ can be rewritten as   
\begin{eqnarray}   
\hat{U}=\sum_{m=0}^\infty\hat{A}_m^\dagger\hat{A}_m,   
\label{U}   
\end{eqnarray}   
where    
\begin{eqnarray}   
\hat{A}_m=2^{-\frac{m}{2}}   
\sum_{k=0}^m\sqrt{{}_m C_k} \ \hat{b}_k\hat{b}_{m-k}   
\label{A}   
\end{eqnarray}   
may be regarded as the annihilation operator of a boson pair with the   
total magnetic quantum number $m$. As we shall consider only the case   
of $g > 0$, the energy minimization for $\hat{U}$ may be exploited to   
estimate the order of magnitude for each term.   
  
The conditions~(\ref{cond1}) allow us to expand $\hat{A}_m$ as   
\begin{eqnarray}   
\hat{A}_0=\hat{b}_0^2, \ \ \ \hat{A}_m=2^{1-\frac{m}{2}}   
\hat{b}_0\hat{b}_m+{\cal O}(N^0) \ \ (m\geq1).   
\label{A1}   
\end{eqnarray}   
Substituting these expansions into Eq.~(\ref{U}),    
we obtain   
\begin{eqnarray}   
\hat{U} = N(N-1) -2N\sum_{m=1}^\infty(1-2^{1-m})\hat{n}_m+{\cal
O}(N^{1/2})    \label{U1}   
\end{eqnarray}   
where the constraint $\hat{N}=N$ is used to eliminate $\hat{n}_0$.        
Because $\hat{n}_2$ and $\hat{n}_3$ appear on the rhs of Eq.~(\ref{U1})   
in a form proportional to $2\hat{n}_2+3\hat{n}_3$, the constraint         
$\hat{L}=L$ may be used to eliminate both of these terms simultaneously.  
We thus obtain   
\begin{eqnarray}   
\hat{U}&=&N\left(N-1-\frac{L}{2}\right) 
\nonumber \\ & &  
+N\left[\frac{\hat{n}_1}{2}+\sum_{m=4}^\infty   
\left(\frac{m}{2}-2+2^{2-m}\right)\hat{n}_m   
\right]+{\cal O}(N^{1/2}). \nonumber \\ & &   
\label{U2}   
\end{eqnarray}   
The minimum of $\hat{U}$ in Eq.~(\ref{U2}) is attained when   
\begin{eqnarray}   
\hat{n}_1={\cal O}(N^{-1/2}), \ \    
\hat{n}_m={\cal O}(N^{-1/2}) \ \ (m\geq4).     
\label{order}   
\end{eqnarray}   
These conditions are compatible with the constraint $\hat{N}=N$ because   
$\hat{n}_0={\cal O}(N^0)$, and with the other constraint $\hat{L}=L$      
because it can be satisfied by an appropriate choice of $\hat{n}_2$ and   
$\hat{n}_3$. We note that the first term in   
Eq.~(\ref{U2}) gives the yrast line~\cite{bert}.   
  
Because the yrast line is dominated by the $m=2$ and $m=3$ modes, the       
low-lying excitations from the line due to binary interactions~(\ref{H4})    
should involve modes with at least up to $m=6$. We therefore use          
$\hat{A}_m$ up to $m=6$ without approximation:   
\begin{eqnarray}   
\hat{A}_0 &=& \hat{b}_0^2 , \ \    
\hat{A}_1=\sqrt{2}\hat{b}_0\hat{b}_1 , \ \    
\hat{A}_2=\hat{b}_0 \hat{b}_2 + \frac{\hat{b}_1^2}{\sqrt{2}} , %\ \    
\nonumber \\ 
\hat{A}_3 &=& \frac{\hat{b}_0\hat{b}_3}{\sqrt{2}}   
+\sqrt{\frac{3}{2}}\hat{b}_1\hat{b}_2 ,  \ \ 
\hat{A}_4 = \frac{\hat{b}_0\hat{b}_4}{2}+\hat{b}_1\hat{b}_3   
+\sqrt{\frac{3}{8}}\hat{b}_2^2,  \nonumber \\    
\hat{A}_5  &=& \frac{\hat{b}_0\hat{b}_5}{\sqrt{8}}   
+\sqrt{\frac{5}{8}}\hat{b}_1\hat{b}_4   
+\frac{\sqrt{5}}{2}\hat{b}_2\hat{b}_3, \nonumber \\    
\hat{A}_6 &=& \frac{\hat{b}_0\hat{b}_6}{4}   
+\sqrt{\frac{3}{8}}\hat{b}_1\hat{b}_5   
+\frac{\sqrt{15}}{4}\hat{b}_2\hat{b}_4 +   
\frac{\sqrt{5}}{4}\hat{b}_3^2.   
\label{A2}   
\end{eqnarray}   
Substituting these expressions into Eq.~(\ref{U}) yields   
\begin{eqnarray}   
\hat{U}&=&N\left(N-1-\frac{L}{2}\right)+   
\frac{27}{34}\hat{n}_3(\hat{n}_3-1)    
+\hat{R}+\hat{S}   
\nonumber \\   
& &    
+\frac{N}{2}\hat{B}_1^\dagger\hat{B}_1   
+\frac{N}{4}\hat{B}_4^\dagger\hat{B}_4   
+\frac{5N}{8}\hat{B}_5^\dagger\hat{B}_5   
+\frac{17N}{16}\hat{B}_6^\dagger\hat{B}_6.  
\label{U30}   
\end{eqnarray}   
Here $\hat{R}$ is given by   
\begin{eqnarray}   
\hat{R}&=&\sum_{k=4}^\infty\left(\frac{k-2}{2}\hat{n}_k^2+\hat{n}_k\right)   
+\sum_{k=6}^\infty\frac{k-3}{2}\hat{n}_1\hat{n}_k  
+\sum_{k=5}^\infty\frac{k-2}{2}\hat{n}_2\hat{n}_k  \nonumber \\ & &  
+\sum_{k=3}^\infty\sum_{l=k+1}^\infty\frac{k+l-4}{2}\hat{n}_k\hat{n}_l   
-2(\hat{n}_3+1)\hat{n}_4-\frac{5}{2}(\hat{n}_4+1)\hat{n}_5   
\nonumber \\   
& &   
-\frac{1}{2}(\sqrt{10}\hat{b}_3^\dagger\hat{b}_4^\dagger\hat{b}_2\hat{b}_5   
+2\sqrt{5}\hat{b}_4^{\dagger 2}\hat{b}_3\hat{b}_5+{\rm h.c.})   
\nonumber \\   
& &   
+\frac{1}{17}   
[23\hat{n}_1\hat{n}_5+32\hat{n}_2\hat{n}_4
\nonumber \\   
& &   
-(14\sqrt{10}\hat{b}_1^\dagger   
\hat{b}_5^\dagger\hat{b}_2\hat{b}_4-\sqrt{30}\hat{b}_1^\dagger\hat{b}_5^\dagger   
\hat{b}_3^2+12\sqrt{3}\hat{b}_2^\dagger\hat{b}_4^\dagger\hat{b}_3^2+   
{\rm h.c.}) ],  \nonumber \\
& &
\label{R}   
\end{eqnarray}   
where h.c. denotes Hermitian conjugates of the preceding terms,  
$\hat{S}$ is given by  
\begin{eqnarray}   
\hat{S}=\sum_{m=7}^\infty\left(\hat{A}_m^\dagger\hat{A}_m   
+\frac{m-4}{2}\hat{n}_0\hat{n}_m\right),   
\label{S}   
\end{eqnarray}   
and $\hat{B}_m \ (m=1,4,5,6)$ are defined by   
\begin{eqnarray}      
\hat{B}_1 &\equiv&  \frac{1}{\sqrt{N}}   
\sum_{k=1}^5\sqrt{k}\hat{b}_{k-1}^\dagger\hat{b}_k, \nonumber \\       
\hat{B}_4 &\equiv& \frac{1}{\sqrt{N}}\left(\hat{b}_0\hat{b}_4   
-2\hat{b}_1\hat{b}_3   
+\sqrt{\frac{3}{2}}\hat{b}_2^2\right), \nonumber \\   
   \hat{B}_5 &\equiv& \frac{1}{\sqrt{N}}\left(   
\hat{b}_0\hat{b}_5-\frac{3}{\sqrt{5}}\hat{b}_1\hat{b}_4   
+\sqrt{\frac{2}{5}}\hat{b}_2\hat{b}_3\right),    
\nonumber \\   
\hat{B}_6 &\equiv& \frac{1}{\sqrt{N}}\left(\hat{b}_0\hat{b}_6   
+\frac{\sqrt{6}}{17}\hat{b}_1\hat{b}_5   
+\frac{\sqrt{15}}{17}\hat{b}_2\hat{b}_4   
+\frac{\sqrt{5}}{17}\hat{b}_3^2\right).   
\label{B}   
\end{eqnarray}   
Note that no approximation is made in obtaining Eq.~(\ref{U30}). It     
follows from~(\ref{order}) that $\hat{R}={\cal O}(N^{-1/2})$. The       
operator $\hat{S}$ is positive semidefinite. To minimize it, we         
consider the states that satisfy   
\begin{eqnarray}   
\hat{b}_k|\Psi\rangle={\rm o}(N^{-1/2})  
\ \ (k\geq7),   
\label{condn}   
\end{eqnarray}   
so as to give $\langle\Psi|\hat{S}|\Psi\rangle={\rm o}(N^0)$, where   
${\rm o}(N^\alpha)$ denotes terms whose order of magnitude is less than   
$N^\alpha$.    
Under the same condition we obtain   
\begin{eqnarray}   
\hat{U}&=&N\left(N-1-\frac{L}{2}\right)+   
\frac{27}{34}\hat{n}_3(\hat{n}_3-1)    
\nonumber \\   
& &    
+\frac{N}{2}\hat{B}_1^\dagger\hat{B}_1   
+\frac{N}{4}\hat{B}_4^\dagger\hat{B}_4   
+\frac{5N}{8}\hat{B}_5^\dagger\hat{B}_5   
\nonumber \\   
& &    
+\frac{17N}{16}\hat{B}_6^\dagger\hat{B}_6+{\rm o}(N^0),  
\label{U3}   
\end{eqnarray}   
which constitutes the primary result of our paper.    
   
Let us define the {\em quasi-degenerate} states, \{$|\Psi\rangle$\},   
as the ones satisfying both condition~(\ref{condn}) and   
\begin{eqnarray}   
& & \hat{B}_m|\Psi\rangle={\rm o}(N^{-\frac{1}{2}})    
\ \ (m=1,4,5,6).   
\label{condB}    
\end{eqnarray}   
The following constraints on these states are then obtained:   
\begin{eqnarray}   
\langle\hat{B}_m^\dagger\hat{B}_m\rangle &=& {\rm o}(N^{-1})    
\ \ (m=1,4,5,6), \label{mini1} \\   
\langle\hat{n}_m\rangle&=&{\rm o}(N^{-1}) \ \ (m\geq7).   
\label{minimum}   
\end{eqnarray}   
The expectation values of $\hat{U}$ for these states are therefore    
given by    
\begin{equation}   
\langle \hat{U} \rangle = N \left (N-1-\frac{L}{2} \right )   
+\frac{27}{34}\,n_3(n_3-1)   
+{\rm o}(N^{0}),   
\label{qdel}   
\end{equation}   
where $n_3$ is the number of octupole excitations, as  
$\hat{Q}_{\lambda}^\dagger \hat{Q}_{\lambda} =    
\hat{n}_{\lambda} +{\cal O}(N^{-1/2})$ for $\lambda \ge 1$.      
Since the population of the octupole mode with  $n_3\geq2$       
raises the energy, while that of the quadrupole mode does   
not\cite{kavou,NU}, the ground-state configuration is given by      
$n_2=L/2$ and $n_3=0$ when $L$ is even and by $n_2=(L-3)/2$      
and $n_3=1$ when $L$ is odd.    
These results are valid up to the order of $N^0$,   
as Eq.~(\ref{qdel}) is valid up to the same order.   
   
As seen in Eq.~(\ref{U3}),  
the quasi-degenerate octupole spectrum lies   
above the yrast states.   
Although the octupole mode couples with other modes through    
\begin{eqnarray}   
[\hat{b}_3,\hat{U}]-\frac{27}{17}\hat{b}_3^\dagger\hat{b}_3^2   
&=& \frac{\sqrt{3N}}{2}\hat{b}_2\hat{B}_1   
+\sqrt{N}\hat{B}_1^\dagger\hat{b}_4   
-\frac{\sqrt{N}}{2}\hat{b}_1^\dagger\hat{B}_4   
\nonumber \\
& &
+\frac{\sqrt{10N}}{8}\hat{b}_2^\dagger\hat{B}_5   
+\frac{\sqrt{5N}}{8}\hat{b}_3^\dagger\hat{B}_6,   
\label{oct}   
\end{eqnarray}   
the expectation value of the rhs over the quasi-degenerate       
state satisfying Eq.~(\ref{condB}) is on the order of ${\rm o}(N^0)$.   
The quasi-degenerate spectrum $27n_3(n_3-1)/34$ is therefore valid   
up to the same order.   
   
The representation of $\hat{U}$ in Eq.~(\ref{U3}) not only describes    
the quasi-degenerate yrast spectra, as discussed above, but also        
explains higher excitations whose energies are linear in $N$.      
Since   
\begin{eqnarray}   
[\hat{B}_m,\hat{B}_n] &=& {\cal O}(N^{-1/2}) \ \ {\rm  and} \ \    
[\hat{B}_m,\hat{B}_n^{\dagger}]=\delta _{m n}+{\cal O}(N^{-1/2}), 
\nonumber \\ 
& & \label{comB}   
\end{eqnarray}   
each $\hat{B}_m$ ($m=1,4,5,6$)   
%(and, of course, $\hat{b}_m$ for $m \geq 7$)   
behaves as a boson-like operator and satisfies   
\begin{eqnarray}   
\hat{B}_m^{\dagger} \hat{B}_m    
= \hat{n}_m + {\cal O}(N^{-1/2}).    
\end{eqnarray}   
Because of the definitions~(\ref{B}) of $\hat{B}_m$ and from~(\ref{order}),  
the excitation energies for $\hat{B}_4$, $\hat{B}_5$, and $\hat{B}_6$ are    
expected to be      
\begin{eqnarray}   
& & \epsilon _0+\epsilon _4 -2\epsilon _2 = gN/4, \nonumber \\   
& & \epsilon _0+\epsilon _5 -\epsilon _2 -\epsilon _3 = 5gN/8,    
\ \ {\rm and}   
\nonumber \\   
& & \epsilon _0+\epsilon _6 -2\epsilon _3 = 17gN/16,    
\end{eqnarray}   
respectively, and the excitation energies for $\hat{B}_1$   
are expected to be    
\begin{eqnarray}   
\epsilon _1+\epsilon _2 -\epsilon _0 -\epsilon _3 = gN/2 \ \ {\rm and} \ \    
2 \epsilon _1-\epsilon _0 -\epsilon _2 = gN   
\end{eqnarray}   
for $n_1=1$ and $n_1=2$, respectively.   
Here    
\begin{eqnarray}   
\epsilon _{\lambda} = - 2gN(1-2^{1-\lambda})   
\end{eqnarray}   
gives the energy of a single excitation of   
$\hat{Q}_\lambda$ ($\lambda \ge 1$)~\cite{mot}   
and $\epsilon _0 \equiv 0$. In fact, these excitation         
energies appear as the coefficients for $\hat{B}_m^\dagger\hat{B}_m$ in   
Eq.~(\ref{U3}).   
 
%************************************************************************ 
 
Here we discuss the physical meaning of Eq.~(\ref{condB}). 
In the BEC the $m=0$ state is macroscopically occupied.  
When the BEC has angular momentum $L$ ($1 \ll L \ll N$),  
the occupation numbers of $m=2$ and $m=3$ states become 
of the order of $L$.  
Because of the interactions, the following scattering processes,  
whose energy contributions are of the order of ${\cal O}(N^0)$,  
become important:  $0+2 \leftrightarrow 1+1$, $0+3 \leftrightarrow 1+2$, 
$0+4 \leftrightarrow 2+2$, $0+5 \leftrightarrow 2+3$, and  
$0+6 \leftrightarrow 3+3$. 
The corresponding terms in the interaction Hamiltonian  
can be diagonalized by the introduction of the quasiparticle 
operators $B_m$ ($m=1,4,5,6$) defined in Eq.~(\ref{B}).  
This procedure is nothing but the renormalization of the bare  
operators $b_m$ due to the interaction.  
The true vacuum is therefore not the bare one defined by  
$b_m |\Psi \rangle = 0$  ($m=1,4,5,6$)  
but that of the quasiparticles defined by $B_m |\Psi \rangle = 0$. 
This is the physical meaning of the constraint (\ref{condB}). 
 
On the other hand, 
the contact interaction dictates that neither quadrupole ($m=2$)  
nor octupole ($m=3$) mode can make ${\cal O}(N^0)$ energy contributions  
together with higher angular momentum states ($m=7$ or higher).  
Hence the supplementary constraint (\ref{condn}) for $m \ge 7$ 
should be imposed for the complete definition of the (interacting)  
vacuum state. In fact, from Eq.~(\ref{S}) $b_m$ excitations are shown 
to need energy costs which are linear in $N$.  
We note that 
because of this renormalization procedure of the quasiparticle operators, 
our excitation spectrum achieves  
the accuracy of ${\rm o}(N^0)$ for quasi-degenerate states. 
Moreover it should be noted that the other low-lying excitations  
can also be explained naturally in terms of $B_m$ and $b_m$ excitations. 
 
%************************************************************************ 

\begin{figure}[t]   
%\vspace*{0.5cm} 
\begin{center} 
\leavevmode\epsfxsize=68mm \epsfbox{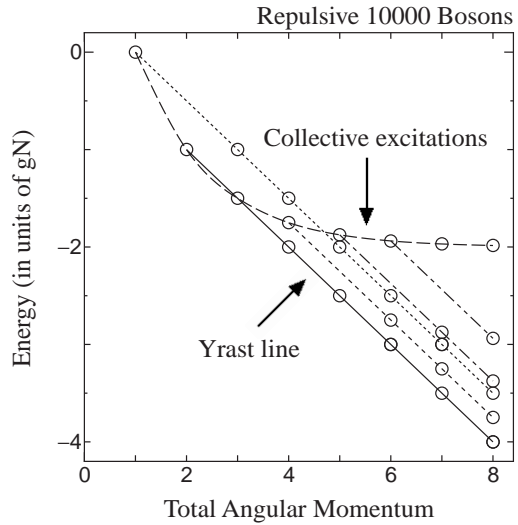} 
%\leavevmode\epsfxsize=136mm \epsfbox{fig1a.eps} 
\end{center} 
%\vspace*{-1.0cm} 
\caption{   
Energies (in units of $gN$) of the yrast states and those of       
low-lying excited states with $N=10000$ for $1 \leq L \leq 8$.     
Open circles show our numerical results, the solid line            
indicates the yrast line $-gNL/2$, and the broken curve connects   
the collective modes excited by $\hat{Q}_\lambda^\dagger$.         
The dashed, dotted, dash-dotted, and dash-double-dotted lines      
correspond, respectively, to the $gN/4$, $gN/2$, $5gN/8$, and      
$17gN/16$ excitations predicted by Eq.~(\protect\ref{U3}).          
}   
\label{fig}   
\end{figure}

As an independent check of the above analytical results, we have          
numerically performed an exact diagonalization of the Hamiltonian,        
confirming that the constraints~(\ref{minimum}) on the quasi-degenerate   
states are satisfied. The excitations with energies linear in $N$ are     
also found to exist and are shown in Fig.~\ref{fig} as open circles.      
Here the solid line indicates the yrast line $-gNL/2$, and the dashed,    
dotted, dash-dotted, and dash-double-dotted lines correspond, respectively,   
to the $gN/4$, $gN/2$, $5gN/8$, and $17gN/16$ excitations that are        
predicted by Eq.~(\ref{U3}).   
The broken curve connects the points of collective excitations   
$\hat{Q}_\lambda^\dagger | 0 \rangle$.   
We see that the excitations shown in Fig.~\ref{fig} can be well explained  
in terms of single excitations of $\hat{B}_m$     
($m=1,4,5,6$).   
When these higher modes are excited, the quasi-degenerate   
conditions~(\ref{condB}) are no longer met,   
so that the coupling of these modes to   
the octupole mode through Eq.~(\ref{oct}) becomes significant.          
This coupling causes ${\cal O}(N^0)$-order corrections to the           
excitation energies, which are on the order of ${\cal O}(N)$.           
The examples of our numerical calculations demonstrating these small    
corrections are shown in Table~\ref{tab}. Since the relative order of   
magnitude of these corrections is on the order of $1/N$, they cannot    
be discerned in Fig.~\ref{fig}.   

Our method provides some constraints on the quasi-degenerate states,      
including the yrast state, through~(\ref{condn})  
and~(\ref{condB}). From these constraints, we may gain some insight  
into quantum fluctuations   
occurring in the quasi-degenerate states \{$|\Psi\rangle$\}.    
For example, from the condition    
$\hat{B}_1|\Psi\rangle ={\rm o}(N^{-\frac{1}{2}})$   
and~(\ref{order}), we obtain    
\begin{eqnarray}    
 \left(    
\hat{b}_0^\dagger\hat{b}_1+\sqrt{2}\,\hat{b}_1^\dagger\hat{b}_2    
+\sqrt{3}\,\hat{b}_2^\dagger\hat{b}_3\right)|\Psi\rangle    
={\rm o}(N^{0}).   
\label{condB1}    
\end{eqnarray}    
This equation strongly suggests that the interconversions between the states   
with $m=0,1,2,3$ (and the interconversions between quadrupole and         
octupole modes~\cite{NU}) play an important role in the quantum           
fluctuations of the quasi-degenerate states. We stress here that the      
rhs of Eq.~(\ref{condB1}) is not ${\cal O}(N^{0})$ but ${\rm o}(N^{0})$.   
Similarly, from the condition   
$\hat{B}_5|\Psi\rangle ={\rm o}(N^{-\frac{1}{2}})$,   
we obtain    
\begin{eqnarray}    
\left(    
\hat{b}_0 \hat{b}_5    
+\frac{\sqrt{10}}{5}\,   
\hat{b}_2\hat{b}_3\right)|\Psi\rangle    
={\rm o}(N^{0}).    
\label{condB5}    
\end{eqnarray}    
It follows from these two constraints that on a mean-field level we have   
$\langle\hat{n}_1\rangle\simeq15\langle\hat{n}_5\rangle/2$. From  
the constraints    
$\hat{B}_m|\Psi\rangle={\rm o}(N^{-\frac{1}{2}})$ ($m=4,5,6$),    
we may also have   
$\langle\hat{n}_4\rangle\simeq3\langle\hat{n}_2\rangle^2/2N$,    
$\langle\hat{n}_5\rangle   
\simeq 2\langle\hat{n}_2\rangle\langle\hat{n}_3\rangle/5N$, and     
$\langle\hat{n}_6\rangle\simeq 5\langle\hat{n}_3\rangle^2/289N$,   
respectively.    
Except for the numerical coefficient in the last relation,    
these results agree with the mean-field analysis in Ref.~\cite{kavou},   
although these approximate relations are not always satisfied by   
exact numerical results.   

%%%%%%%%%%%%%%%%%%%%%%%   
\section{Conclusions}     
\label{sec:Conclusions}   
   
In this paper we have proposed an operator-algebraic approach to analyzing    
the yrast spectrum of weakly interacting bosons. The basic idea is to    
exploit the macroscopic occupation of a particular mode to rewrite the   
second-quantized interaction Hamiltonian under the restriction of the    
total particle-number conservation and the angular-momentum conservation.   
   
We have analytically shown that the lowest-lying excitation  
spectrum from the yrast line is given by $27gn_3(n_3-1)/34$.    
Equation~(\ref{U3}) also shows that this quasi-degenerate energy         
spectrum is separated from other levels by energy gaps that are   
linear in $N$. These results are in agreement with what we have found by  
numerical diagonalization of the same Hamiltonian~\cite{NU}.             
Equation~(\ref{U3}) can also explain the excitations whose energies      
are linear in $N$, again in agreement with our exact diagonalization     
calculations~\cite{NU}.   
   
\vspace*{0.45cm}

The method presented herein only exploits the presence of      
a macroscopically occupied state, and it should therefore not be      
restricted to the case of $L/N\ll1$ discussed above. Such an     
extension is in progress and will be reported elsewhere.   

\begin{acknowledgments}   
   
M.U. acknowledges support by a Grant-in-Aid for Scientific Research   
(Grant No. 11216204) by the Ministry of Education, Science, Sports,   
and Culture of Japan, and by the Toray Science Foundation.            
T.N. was supported for research in the USA by the Japan Society   
for the Promotion of Science.   
   
\end{acknowledgments}

\bigskip   
\noindent   
{\it Note added:}    
After completion of this work, we became aware of three        
studies (cond-mat/0011303, cond-mat/00011430, and cond-mat/00012438)    
that used different methods to derive the same analytic formula    
$27gn_3(n_3-1)/34$, as presented herein.

\end{multicols} 
\widetext  
   
\begin{table}   
\begin{tabular}[t]{||c|c|c|c|c|c|c|c||}    
$L$ & 7 & 7 & 7 & 8 & 8 & 8 & 8 \\   
\hline   
excitation energy   
& 2502.8   
& 5000.0   
& 6253.3   
& 2507.7   
& 5000.0   
& 6254.0   
& 10628   
\end{tabular}   
   
\bigskip   
\caption{Excitation energies measured from the yrast line with    
$N=10000$ for $L=7$ and 8.   
The numerical results agree with our analytical predictions of    
$N/4=2500$, $N/2=5000$, $5N/8=6250$, and $17N/16=10625$ with      
${\cal O}(N^{0})$ deviations that arise from the coupling to     
the octupole mode through Eq.~(\protect\ref{oct}).   
}   
\label{tab}   
\end{table}   
      
\end{document}